\documentclass[twocolumn,showpacs,preprintnumbers,amsmath,amssymb]{revtex4}

\usepackage{graphicx}
\usepackage{dcolumn}
\usepackage{bm}

\begin{document}


\title{Semi-Analytic Estimates of Lyapunov Exponents in Lower-Dimensional
Systems}

\author{Bal{\v s}a Terzi{\'c}}
 \altaffiliation{Corresponding author. \\ 
{\it Email addresses:} bterzic@astro.ufl.edu, kandrup@astro.ufl.edu}
\affiliation{Department of Astronomy, University of Florida, Gainesville, Florida 32611}%

\author{Henry Kandrup}
 \affiliation{Department of Astronomy, Department of Physics, and 
  Institute for Fundamental Theory \\
  University of Florida, Gainesville, Florida 32611}

\date{\today}

\begin{abstract}
Statistical arguments, seemingly well-justified in higher dimensions, can also 
be used to derive reasonable estimates of Lyapunov exponents $\chi$ in 
lower-dimensional Hamiltonian systems. This letter explores the assumptions 
incorporated into these arguments. The predicted $\chi$'s are insensitive to 
most details, but do depend sensitively on the nongeneric form of the 
auto-correlation function characterising the time-dependence of an orbit. 
This dependence on dynamics implies a fundamental limitation to the 
application of thermodynamic arguments to lower-dimensional systems.
\end{abstract}

\maketitle

\section{\label{sec:intro}Introduction}
Computing Lyapunov exponents, which measure the average linear stability
of chaotic orbits in a $t\to\infty$ limit, entails solving a matrix harmonic
oscillator equation with time-dependent frequencies. 
The differential-geometric analogue is a Jacobi equation (equation of geodesic
deviation) on a manifold equipped with the Eisenhart metric \cite{e29}.
To the extent that aperiodic chaotic orbits can be treated as `random,' this 
exact equation can be approximated as a stochastic harmonic oscillator 
equation with `randomly varying' frequencies \cite{p93}.
The implementation of such a picture, along with more detailed assumptions 
that would seem reasonable in a thermodynamic limit, has yielded remarkable
success in estimating Lyapunov exponents in high-dimensional systems 
\cite{ccp96}.
In lower dimensions, where some of these assumptions are at least questionable,
one still derives estimates that are qualitatively correct; but while the
general trends of the dynamics tend to be captured, many details are often
missed \cite{ksb02,cp02}.

The aim of this letter is two-fold, namely (i) to describe how and to what
extent this general approach can be improved upon by relaxing some of the
underlying assumptions, and (ii) to determine physically why, in
lower dimensions, much of the dynamics eludes a statistical description.
The analysis will focus on Hamiltonians of the form
$H=\sum_{i=1}^{D}p_{i}^{2}/2+V(\{q_{i}\})$ for two 
representative potentials, namely the Fermi-Pasta-Ulam (FPU) potential,
\begin{equation} \label{FPU}
V(q_1,q_2,...,q_D) = \sum\limits_{i=1}^{D}
\left[ {a\over 2} (q_{i+1} - q_i)^2 + {b\over 4} (q_{i+1} - q_i)^4
    \right],
\end{equation}
and a higher-dimensional analogue of the dihedral potential,
\begin{equation} \label{dih}
V(q_1,q_2,...,q_D) = - \sum\limits_{i=1}^{D} q_i^2 +
{1\over 4} \left( \sum\limits_{i=1}^{D} q_i^2 \right)^2 -
{1\over 4} \sum\limits_{i<j=1}^{D} q_i^2 q_j^2.
\end{equation}

\section{\label{sec:tech}Relaxing underlying assumptions}
The key physical assumption underlying the entire analysis is that the exact 
Jacobi equation can be approximated by a matrix stochastic oscillator equation 
in which the curvature, which determines the frequencies, varies `randomly'
along a chaotic orbit, {\em i.e.,} that the `true' dynamical equation can
be replaced by an approximate statistical equation. Given this assumption, 
what remains is to specify the statistics of the random process. Basically
this has involved three additional assumptions \cite{ccp96}:
\par\noindent (1) An assumption of {\em isotropy}, which reduces the 
$D$-dimensional matrix equation to a one-dimensional oscillator equation.
\par\noindent (2) An assumption that the {\em curvature fluctuations acting
at any given instant are Gaussian-distributed.}
\par\noindent (3) An assumption regarding the form of the {\em auto-correlation
function} and an associated {\em correlation time} ${\tau}_{c}$, which 
characterise changes in the curvature fluctuations along a chaotic orbit.

Mathematically, the reduction from a $D$-dimensional matrix equation to a
one-dimensional stochastic oscillator follows from the assumption that the
configuration space can be approximated as quasi-isotropic, {\em i.e.,} that
if the Riemann tensor be viewed as the sum of a Ricci curvature $K$ and a
non-isotropic Weyl projective tensor, the latter may be assumed to vanish. 
Physically, this is equivalent to assuming that there are no preferred 
directions in space, so that the matrix equation reduces to $D$ identical
one-dimensional equations. At least for potentials in which different degrees
of freedom enter identically, it might seem that this assumption becomes 
increasingly justified in higher dimensions. Earlier investigations found
that, for some potentials, the analytic estimate ${\chi}_{ana}$ converges
quite rapidly towards the value ${\chi}_{num}$ computed numerically, but that
for other potentials it does not \cite{ksb02}. This suggests that 
quasi-isotropy is likely {\em not} the principal source of error.

It remains, therefore, to consider the form of the curvature fluctuations
both at a fixed instant and as a function of time. 
To the extent that the entire phase space is chaotic, the form of the 
curvature fluctuations at a fixed instant can be addressed either 
analytically, given an assumption of ergodicity, or numerically, by evaluating
the statistical properties of representative orbit ensembles. However, it 
would appear that dynamical issues involving the auto-correlation function and 
the correlation time must be addressed numerically.

\subsection{\label{ssec:gauss}Approximating curvature fluctuation as 
Gaussian-distributed}
The assumption is that curvature fluctuations in different directions can be
approximated as nearly independent and, at any given instant, 
Gaussian-distributed. For potentials investigated in \cite{ksb02}, this 
approximation was completely unjustified in systems with two degrees of 
freedom (dof), and questionable in slightly higher-dimensional systems. 
To determine the accuracy of such an approximation, we compare the Gaussian
constructed from numerical estimates of the first two moments of the curvature
to the exact curvature distribution function $N[K]$, which can be computed
analytically for potentials in which the radius $r$ can be expressed in terms
of the Ricci curvature $K$. 

\begin{figure*}
\begin{center}
\includegraphics[height=12cm]{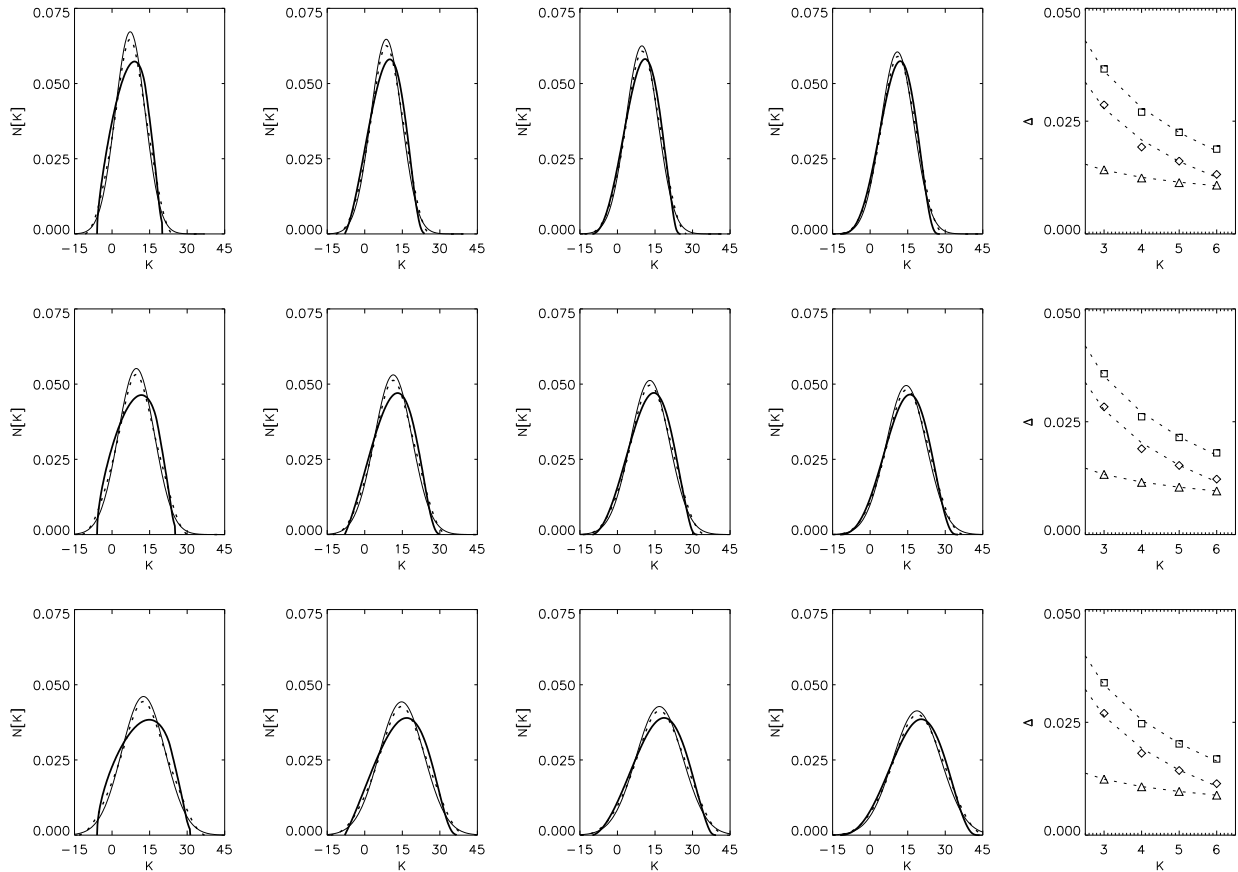}
\caption{
The distribution of curvatures, $N[K]$, approximated by a Gaussian $Z(K)$ 
(thin solid line), cumulant expansion $C_{12}(K)$ (dashed line) and exact 
(thick solid line), as computed by eq. (\ref{N_KN}), for the dihedral potential
with dimensionality $D = 3, 4, 5, 6$ and energy levels $E = 1, 3, 6$.  
The last column shows the absolute difference between the exact value of 
$N[K]$ and the purely Gaussian approximation $Z(K)$ (squares),
exact value and the cumulant expansion $C_{12}(K)$ (diamonds) and 
the Gaussian approximation and the cumulant expansion (triangles).
A function of the form $(a/\sqrt{D}) + b$ is fit through data points 
(dashed lines).  
}
\label{fig:1}
\end{center}
\end{figure*}

Given the assumption of a microcanonical distribution, moments of $K$ satisfy
\begin{equation} \label{avg_KK2}
\langle K^n \rangle = {{\int\limits_{V({\bf q})=E} d {\bf q} \
[E-V({\bf q})]^{(D-2)/2} ~ K^n({\bf q})}\over {\int\limits_{V({\bf
q})=E} d {\bf q} \ [E-V({\bf q})]^{(D-2)/2}}},
\end{equation}
where $D$ is the dimensionality. 
If one can transform from the coordinates $(q_{1},q_{2},...,q_{D})$ to 
$(K,{\theta}_{1},{\theta}_{2},...,{\theta}_{D-1})$, where $K$ labels surfaces
of constant curvature (density), eq.~(\ref{avg_KK2}) becomes

\begin{equation} \label{avg_K_new}
\langle K^n \rangle = { {\int K^n dK \int d\theta_1 d\theta_2 ...
d\theta_{D-1} 
\left| {{\partial {\bf q}}\over {\partial (K,{\bf \theta)}}} \right|
\left[E-V\right]^{{D-2}\over 2}
} \over {\int dK \int d\theta_1 d\theta_2 ... d\theta_{D-1}
\left| {{\partial {\bf q}}\over {\partial (K,{\bf \theta)}}} \right|
\left[E-V\right]^{{D-2}\over 2}
}}
\end{equation}
and the distribution function
\begin{equation} \label{N_K}
N[K] = \int d\theta_1 d\theta_2 ... d\theta_{D-1} \left|
{{\partial {\bf q}}\over {\partial (K,{\bf \theta)}}} \right|
\left[E-V\right]^{{D-2}\over 2},
\end{equation}
where the dependence of 
$\left| {{\partial {\bf q}}\over {\partial (K,{\bf \theta)}}} \right|$ 
and $V$ on the transformed coordinates is omitted for brevity. 
For compact isoenergy surfaces, the integration extends over the entire 
range of angles $\theta_i$.
For a $D$-dimensional potential, a natural choice for the $\theta_i$ are the 
angles $\psi_i$ of generalized polar coordinates $(r,\psi_1,...,\psi_{D-1})$, 
with $0 < \psi_1,\psi_2,..., \psi_{D-2} < \pi$ and $0 < \psi_{D-1} < 2 \pi$,
in terms of which the distribution of curvatures becomes
\begin{eqnarray} \label{N_KN}
N[K]&=&\int\limits^{2 \pi}_{0} d \psi_{D-1} \int\limits^{\pi}_{0}
d\psi_{D-2} ... \int\limits^{\pi}_{0} d\psi_1 
{{\partial r}\over {\partial K}} \left[E-V\right]^{{D-2}\over 2} 
\nonumber \\
&\times& r^{D-1} \sin^{D-2} \psi_1 \sin^{D-3} \psi_2 ... \sin \psi_{D-2}.
\end{eqnarray}
The results of the direct numerical evaluation of this $(D-1)$-dimensional
integral for the dihedral potential are shown in Fig. \ref{fig:1} (thick lines).

If the shape of the curvature distribution does not deviate significantly 
from a Gaussian, {\em i.e.}, if the higher moments of the curvature are 
non-vanishing but small compared to unity, a better estimate can be obtained 
through the asymptotic cumulant expansion \cite{as74}
\begin{eqnarray} \label{Cas}
C_m (x) &=& {\bar Z}(x) - \left[{\gamma_1 \over 6} {\bar Z}(x)^{(3)}\right] \\
      &+& \left[{\gamma_2 \over 24} {\bar Z}(x)^{(4)} +
                {\gamma_1^2 \over 72} {\bar Z}(x)^{(6)} \right] \nonumber \\
      &-& \left[{\gamma_1 \gamma_2 \over 144} {\bar Z}(x)^{(7)} +
                {\gamma_1^3 \over 1296} {\bar Z}(x)^{(9)} \right] \nonumber \\
      &+& \left[{\gamma_2^2 \over 1152} {\bar Z}(x)^{(8)} +
                {\gamma_1^2 \gamma_2 \over 1728} {\bar Z}(x)^{(10)} +
                {\gamma_1^4 \over 31104} {\bar Z}(x)^{(12)} \right] \nonumber \\
      &+& ... \  \,{\equiv}\; p_m(x)  Z(x),
\nonumber
\end{eqnarray}
where
\begin{eqnarray} \label{gammas}
{\bar Z}(x) & = & {1\over{\sqrt{2 \pi}}} e^{-{{x^2}\over{2}}}, \\
Z (x) & = & {1\over{\sigma}} {\bar Z}\left({{x-\langle K \rangle}
\over{\sigma}}\right), \nonumber \\
\mu_n & = & \left< \left( K- \left< K \right> \right)^n \right>, \nonumber \\
\sigma & = & \sqrt{\mu_2}, \nonumber \\
\gamma_1 & = & {\mu_3 \over \sigma^3}, \nonumber \\
\gamma_2 & = & {\mu_4 \over \sigma^4} - 3. \nonumber \\
\end{eqnarray}
$p_m (x)$ is a polynomial of degree $m$, where $m$ is the highest derivative 
of $Z(x)$ kept in the expansion (\ref{Cas}).  $\gamma_1$ is the skewness
coefficient and $\gamma_2$ the kurtosis. 

Fig. \ref{fig:1} shows three representations $N_k(K)$ of the curvature 
distribution for the dihedral potential: the Gaussian approximation, 
the cumulant expansion, and the exact value computed from the 
eq. (\ref{N_KN}), along with the difference between them as measured by 
$\Delta(N_i,N_j)=
\left[{\int\limits^{\infty}_{-\infty}\left(N_i(K)-N_j(K)\right)^2 dK}
\right]^{1/2}$.
Given that all the approximations are normalized, {\em i.e.} that they 
enclose unit surface, $\Delta$ represents the relative difference between the 
approximations.  It is apparent that, as the dimensionality $D$ is increased, 
the convergence of the exact distribution towards a Gaussian distribution is 
of order $1/\sqrt{D}$, which is the usual thermodynamical behavior.
The cumulant expansion provides slightly more accurate estimates to the 
curvature distribution than the Gaussian approximation, but also converges as 
$1/\sqrt{D}$.

\subsection{\label{sec:acf}Auto-correlation function 
$\Gamma (t) \equiv {\langle}K(0)K(t){\rangle}$}
\begin{figure*}
\begin{center}
\includegraphics[width=6in]{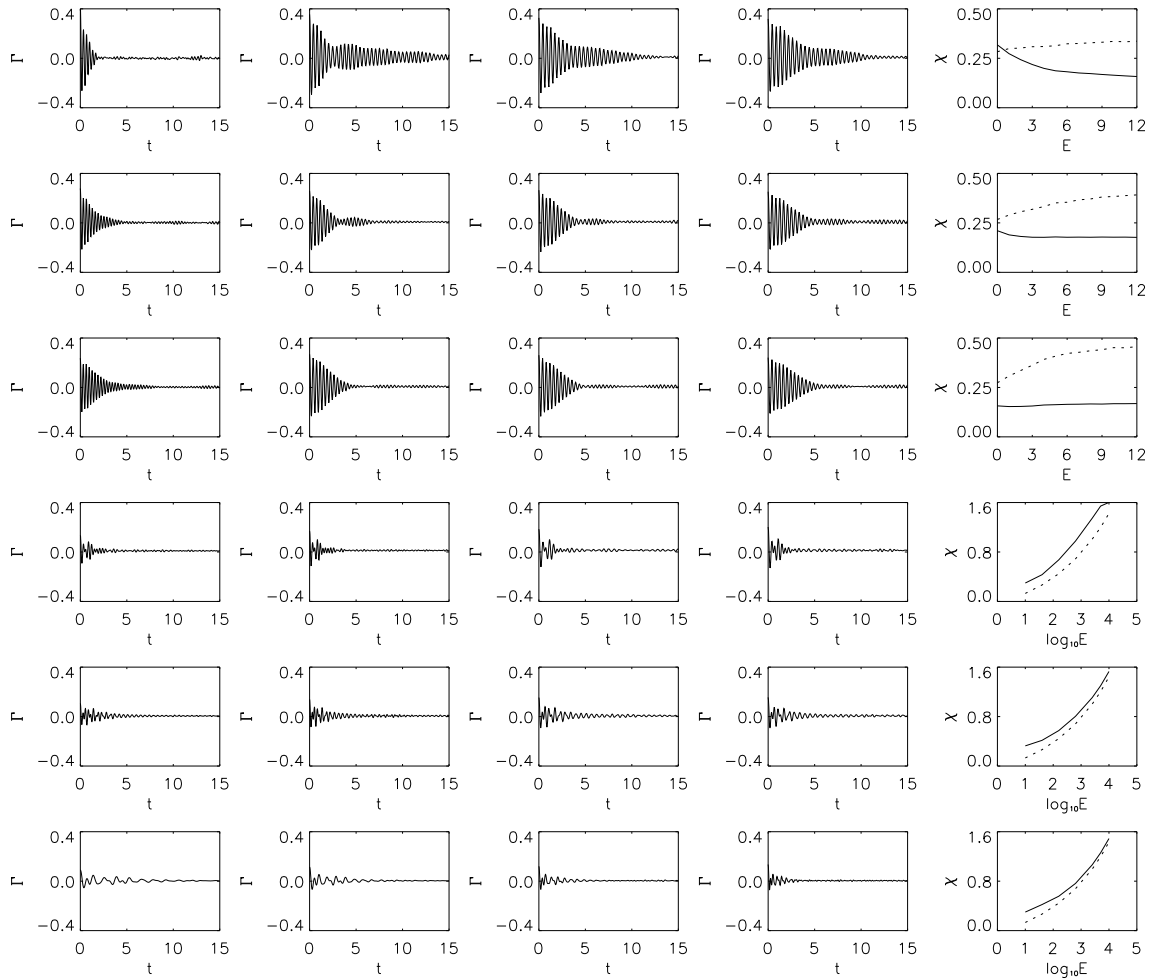}
\caption{
Auto-correlation function $\Gamma$ in units of dynamical times for the $3$-, 
$6$- and $9$-dimensional dihedral potential with energies $E = 0, 4, 8, 12$, 
(top three rows, respectively), and $4$-, $5$- and $6$-dimensional FPU 
potential with energies $\log_{10} E = 1, 1.6, 2.8, 3.4$ (bottom three rows,
respectively), along with the analytic (dashed lines) and numerical estimates 
(solid lines) of the Lyapunov exponents (rightmost column).
}
\label{fig:2}
\end{center}
\end{figure*}

The stochastic oscillator equation can be solved analytically using a 
method developed by van Kampen \cite{vk76} if one assumes that 
the correlation time ${\tau}_{c}$, defined such that 
the auto-correlation function
${\Gamma}(t)\equiv \langle K(0) K(t)\rangle\;{\approx}\;0$ for $t>{\tau}_{c}$, 
is short compared with the 
duration of the stochastic process.  When the correlation time
scaled in terms of the the average orbital time at that energy is short,
typically less than ${\bar t}=2$, which is the case for the $3$-dimensional
dihedral potential at low energies and the FPU potential for all 
dimensions and energies, accurate estimates of the Lyapunov exponents are 
found (Fig. \ref{fig:2}).  It is also evident that, for the dihedral 
potential, the correlation time grows with energy, which renders van Kampen's 
analytic solution increasingly inaccurate.  Because of the long tail of the
auto-correlation function, even in higher dimensions the agreement between 
the numerical results and the geometric estimates does not improve.
Empirically, it appears that the threshold at which van Kampen's 
method yields accurate estimates is ${\tau}_{c} \equiv {\bar t}/2 
\lessapprox 1$.  This is consistent with the fact that van Kampen's solution 
is the first term in a power series expansion in $\tau_c$,
with error $\propto{\;}{\tau}_{c}^2$. 

In point of fact, estimating Lyapunov exponents as outlined in 
\cite{p93,ccp96,ksb02} assumes that the stochastic process is 
delta-correlated, {\em i.e.,} $\Gamma(t) = \tau \sigma^2 \delta(t)$, 
where ${\tau}$ represents a characteristic time scale. However, that 
assumption was introduced purely on grounds of computational convenience, and 
does {\em not} accurately reflect the shape of the actual correlation 
function.  Rather, analysis of data such as those exhibited in Fig. 
\ref{fig:2}, for both the dihedral and FPU potentials, reveals that the 
auto-correlation function is well fit by the functional form
\begin{equation} \label{acf}
\Gamma(t) = A e^{-p t} \sin \omega t,
\end{equation}
where ${\omega}^{-1}$ is comparable to, and strongly correlates with, a 
typical orbital time scale (Fig. \ref{fig:3}). The damping rate $p$ appears to 
correlate with the rate ${\lambda}$ at which an initially localized ensemble 
of orbits would evolve towards a microcanonical equilibrium \cite{kn02}, 
which implies that it is comparable to, but somewhat smaller than, the value 
of the largest Lyapunov exponent -- typically 
${\lambda}{\;}{\sim}{\;}0.2-0.65{\chi}$.

\subsection{\label{sec:act}Characteristic time scale of the process $\tau$}
\begin{figure*}
\begin{center}
\includegraphics[width=6in]{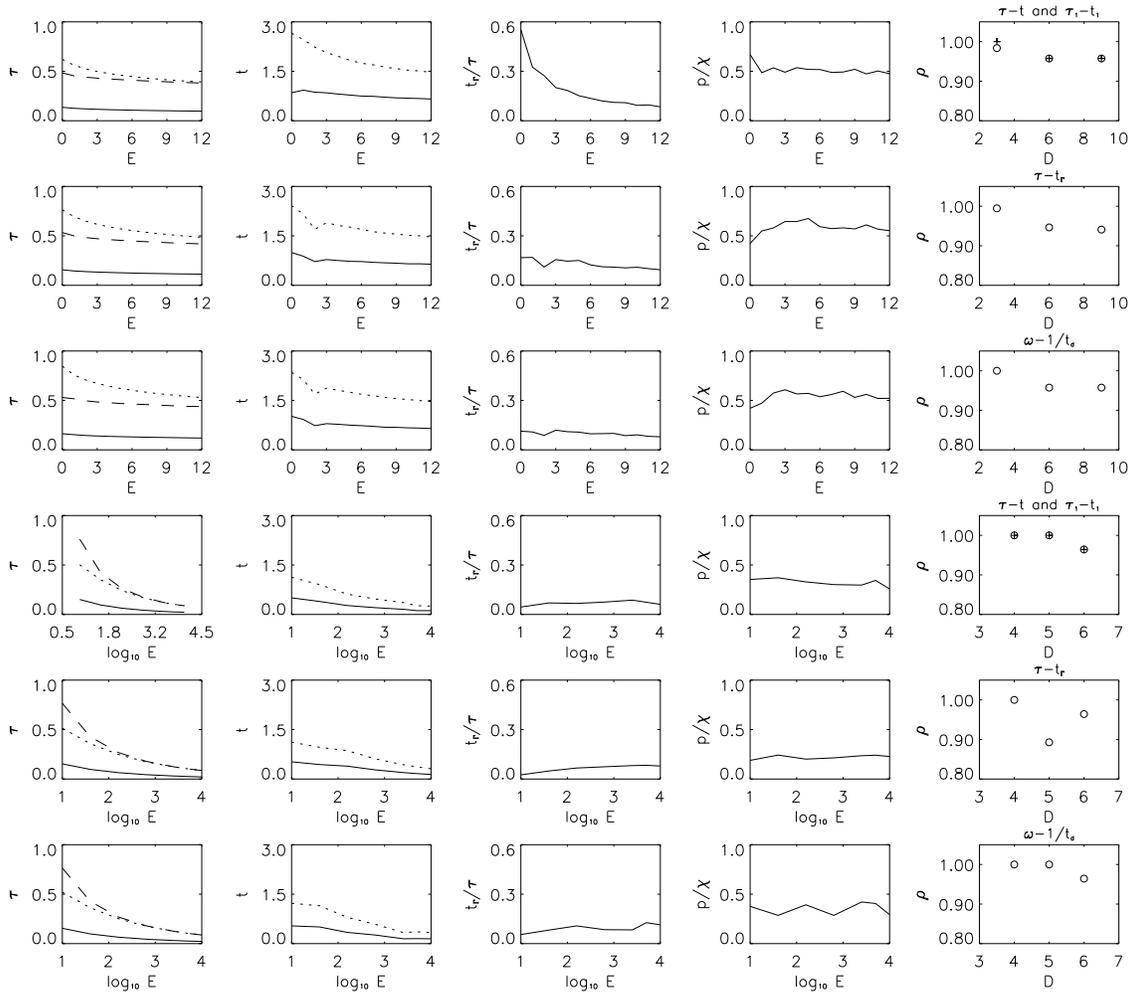}
\caption{
Top three rows represent the $3$-, $6$- and $9$-dimensional dihedral 
potential, and bottom three the $4$-, $5$- and $6$-dimensional FPU potential. 
First column: geometric characteristic times $\tau$ (solid lines), 
$\tau_1$ (short dashes) and $\tau_2$ (long dashes).  Second column:
numerically obtained characteristic times $t_1$ (solid) and $t$ (dashed lines), 
as defined in (\ref{ts}).  Third column: characteristic time estimate 
$t_{\Gamma}$ scaled to $\tau$.  Fourth column: ratio of the damping rate 
$p$ and the largest Lyapunov exponents $\chi$.  Fifth column: for the 
dihedral and FPU potentials as a function of dimensionality, the Spearman 
correlation between $\tau$ and $t$ (crosses) and $\tau_1$ and $t_1$ 
(circles) (first and fourth panel), $\tau$ and $t_{\Gamma}$ (second and fifth 
panel) and $\omega$ and the dynamical time $t_d$ (third and sixth panel).
}
\label{fig:3}
\end{center}
\end{figure*}
In earlier work, the characteristic time scale ${\tau}$
was defined in a seemingly {\em ad hoc} fashion as the geometric mean of
two time scales which were identified using purely geometric considerations,
{\em i.e.,} ${\tau}^{-1}=2({\tau}_{1}^{-1}+{\tau}_{2}^{-1})$, where
${\tau}_{1}$ represents the typical time between successive conjugate points
on the manifold, {\em i.e.,} points where the Jacobi field of geodesic 
deviation vanishes, and ${\tau}_{2}$ corresponds to the length scale on which 
curvature fluctuations become comparable to the average curvature.
In point of fact, however, ${\tau}_{1}$ and ${\tau}_{2}$ are closely related
to the physically relevant time scales entering into the problem: quite
generally, it appears that, to a fair degree of approximation, 
${\tau}_{1}{\;}{\propto}{\;}{\omega}^{-1}$ and, as such, scales as a 
characteristic orbital time scale, and that 
${\tau}_{2}{\;}{\propto}{\;}p^{-1}$ and, as such, scales as the damping time
or, equivalently, the time scale associated with chaotic phase mixing as 
an ensemble evolves towards a time-independent equilibrium \cite{kn02}.  If
we define these two physical time scales and their geometric mean as
\begin{equation}\label{ts}
t_1 = {{2 \pi}\over{\omega}}, \hskip20pt t_2 = {1\over{p}}, \hskip20pt
{t}^{-1}=2({t_1}^{-1}+{t_2}^{-1}),
\end{equation}
we find a strong correlation between $\tau$ and $t$, as well as between
$\tau_1$ and $t_1$ (first and fourth panel in the rightmost column of
Fig. \ref{fig:3}).  The time $t_2$ is typically an order of magnitude
longer than $t_1$, which means that the characteristic time $t$ is 
determined primarily by the shorter time scale $t_1$, the orbital time scale.

Assuming proportionality constants that are comparable for both scalings
the geometric mean would thus involve ${\tau}{\;}{\propto}{\;}1/(p+{\omega})$.
In point of fact, an auto-correlation function of the form (\ref{acf}) 
implies a characteristic time scale
\begin{equation} \label{ct}
t_{\Gamma} = {{
\int\limits_0^{\infty} \Gamma(t) t dt
}\over{
\int\limits_0^{\infty} \Gamma(t) dt
}} = {{2 p}\over{p^2 + \omega^2}}.
\end{equation}

Although this differs from the preceding formula for ${\tau}$,
we find an excellent correlation between the time $t_{\Gamma}$ and the 
time scale ${\tau}$ defined geometrically (second and fifth row in the
rightmost column of Fig. \ref{fig:3}).

\section{\label{sec:dynamics}Dynamics}
Statistical estimates of the largest Lyapunov exponent in lower-dimensional
Hamiltonian system entail two types of approximations: (1) 
{\em time-independent} -- the space is approximated as isotropic and the
curvature fluctuations as Gaussian-distribution, and (ii) {\em time-dependent}
-- stochastic oscillations are approximated as a delta-correlated process
with a characteristic time scale ${\tau}$ derived using {\em ad hoc} geometric
arguments. 
The time-independent approximations reduce the dynamics to a one-dimensional
stochastic oscillator equation; the time-dependent approximations are used 
to solve that equation via van Kampen's method.

At least for the two potentials considered here, the time-independent
approximations appear well-justified and are unlikely sources of significant
error. However, numerical computations show that the time-dependent 
approximations are not always justified. 
The simplifications which reduce the exact integral equation for $K(t)$ to
a linear first order differential equation for ${\langle}K(t){\rangle}$
rely entirely on the assumption that the auto-correlation time $t_{c}$ is short
when compared with the duration of the process. 
As is evident from the slow decay of the auto-correlation function, this 
assumption typically fails for the dihedral potential, so that van Kampen's
method is strictly speaking inapplicable. 
The FPU potential has a more rapidly decaying auto-correlation function
with a much shorter auto-correlation time. In this case, van Kampen's method
is applicable and, as illustrated in Fig. \ref{fig:2}, yields more accurate 
estimates of the largest Lyapunov exponents.
That this approach works much better for the FPU potential than for
the dihedral potential may reflect the fact that the FPU system is sparsely
coupled in the sense that each coordinate `interacts' only with the 
neighbouring ones, while in the dihedral potential every coordinate is
coupled to every other one.  This possibility would seem related to the fact
that many properties of dynamical systems appear to depend not simply on 
he number of degrees of freedom, but on their connectance \cite{f78}, a 
measure of the extent to which these degrees of freedom are directly coupled.

Justifying the time-independent approximations reinforces the idea that
chaos in lower-dimensional systems can be modeled by a one-dimensional
stochastic oscillator. However, only in some cases, when the auto-correlation
function damps sufficiently quickly, can van Kampen's method be employed
to estimate solutions to this stochastic oscillator equation.

\vskip .15in
This research was supported in part by AST-0070809.  We would like to 
thank the Florida State University School of Computational Science and 
Information Technology for granting access to their supercomputer facilities.

\bibliography{tk2003}
\end{document}